\documentclass[twocolumn,preprintnumbers,nofootinbib,amsmath,amssymb,amsfonts,aps,prd,balancelastpage,superscriptaddress]{revtex4-2}

\usepackage{color}
\usepackage{mathtools}
\usepackage{amsthm,amstext,amscd}
\usepackage{epsfig,graphicx,bm}
\usepackage{epstopdf, epsf}
\usepackage{dcolumn}
\usepackage{tensor}

\usepackage{datetime}
\newdateformat{monthyeardate}{%
  \monthname[\THEMONTH], \THEYEAR}
  
\usepackage{graphicx,caption,subcaption}
\usepackage[colorlinks,
           linkcolor = blue,		
           citecolor = blue,
           ]{hyperref}

\newcommand{\be}{\begin{equation}}
\newcommand{\ee}{\end{equation}}

\newcommand{\bse}{\begin{subequations}}
\newcommand{\ese}{\end{subequations}}
\newcommand{\bea}{\begin{eqnarray}}
\newcommand{\eea}{\end{eqnarray}}
\newcommand{\ba}{\begin{array}}
\newcommand{\ea}{\end{array}}
\newcommand{\bc}{\begin{center}}
\newcommand{\ec}{\end{center}}

\def\cl{{\cal L}}

\def\co{{\cal O}}

\def\ba{{\bf A}}

\def\bc{{\bf C}}

\def\be{{\bf E}}

\begin{document}

\preprint{IPMU25-0021}

\title{Curvature corrections to Starobinsky inflation can explain the ACT results}

\author{Andrea Addazi}
\email{addazi@scu.edu.cn}
\affiliation{Center for Theoretical Physics, College of Physics Science and Technology, Sichuan University, 610065 Chengdu, China}
\affiliation{Laboratori Nazionali di Frascati INFN, Frascati (Rome), Italy, EU}

\author{Yermek Aldabergenov}
\email{ayermek@fudan.edu.cn}
\affiliation{Department of Physics, Fudan University, 220 Handan Road, Shanghai 200433, China}

\author{Sergei V. Ketov}
\email{ketov@tmu.ac.jp}
\affiliation{Department of Physics, Tokyo Metropolitan University, Tokyo 192-0397, Japan}
\affiliation{Kavli Institute for the Physics and Mathematics of the Universe (WPI),\\
 The University of Tokyo Institutes for Advanced Study, Chiba 277-8583, Japan}
 
\date{\monthyeardate\today}

\begin{abstract}
\noindent
We investigate the impact of curvature corrections to Starobinsky inflation in light of the latest observational results from the Atacama Cosmology Telescope (ACT). While the pure Starobinsky model remains a compelling candidate for cosmic inflation, we explore how the higher-order curvature terms $R^3$, $R^{4}$ and $R^{3/2}$ modify the inflationary predictions. Using the scalar-tensor formulation of $f(R)$ gravity, we derive the effective scalar potentials and compute the resulting scalar tilt $n_s$, its running index $\alpha_s$ and the tensor-to-scalar ratio $r$. We show that those curvature corrections can shift the predictions to align better with the ACT data, thus providing a possible resolution to a minor discrepancy between the standard Starobinsky model and ACT observations. Our findings suggest that the modified Starobinsky models with the higher-order curvature terms offer a viable pathway to reconciling inflationary predictions with precision cosmological measurements. At the same time, measuring or constraining primordial tensor modes can help to discriminate between these corrections.
\end{abstract}

\maketitle

\section{Introduction}

The Starobinsky model of inflation \cite{Starobinsky:1980te}, based on $R + R^2$ gravity, has long been regarded as one of the most successful frameworks for explaining the early universe's accelerated expansion. Its predictions for the scalar spectral index $n_s$ and tensor-to-scalar ratio $r$ are in excellent agreement with \textit{Planck} and \textit{BICEP/Keck} results, making it a leading candidate for describing cosmic inflation, see \cite{Ketov:2025nkr} for a recent review. However, recent measurements from the \textit{Atacama Cosmology Telescope (ACT)}~\cite{ACT:2025fju,ACT:2025tim} 
and other high-precision experiments have introduced subtle deviations that may hint at new physics beyond the simplest Starobinsky model, see the recent literature in Refs.~\cite{Yogesh:2025wak,Chakraborty:2025yms,Haque:2025uis,Herbermann:2025uqz,Gialamas:2025ofz,Liu:2025qca,Efstathiou:2025tie,Yin:2025rrs,Cheng:2025lod,Ye:2025ark,Preston:2025tyl,Zharov:2025evb,Haque:2025uri,Chen:2025mlf,Wang:2025zri,Scherer:2025esj,Fabbian:2025fdk,Drees:2025ngb,ACT:2025rvn,Pallis:2025epn,DESI:2025gwf,Aoki:2025bmj,Lin:2025gne,Specogna:2025guo,Wang:2025owe,He:2025bli,deSouza:2025rhv,Wang:2025bkk,Mirpoorian:2025rfp,Gao:2025onc,Dioguardi:2025mpp,Lonappan:2025hwz,Kim:2025dyi,Antoniadis:2025pfa,Salvio:2025izr,Lee:2025yah,Naokawa:2025shr,Barenboim:2025vrc,Buoninfante:2025dgy,Gialamas:2025kef,Calderon:2025xod,Brahma:2025dio,Dioguardi:2025vci,Berera:2025vsu,Aoki:2025wld,DiValentino:2025sru,Newman:2025noj,Chaussidon:2025npr,Sailer:2025rks,Silva:2025hxw,LiteBIRD:2025mvy,Pan:2025psn,Smith:2025zsg,Kallosh:2025rni,Lee:2025yvn,Nakagawa:2025ejs,Brandenberger:2025hof,Luchina:2025prh,Peng:2025bws}  for more details. When the new \textit{ACT} observations are combined with \textit{Planck} data and baryon acoustic oscillation (BAO) data from the Dark Energy Spectroscopic Instrument (\textit{DESI})~\cite{DESI:2024uvr} denoted as \textit{PACT-LB}, Ref.~\cite{ACT:2025fju} obtains
$n_s = 0.9743 \pm 0.0034$.

In this paper, we study whether the higher-order (perturbative) curvature corrections, such as $R^3$ and $R^4$, and even fractional powers as $R^{3/2}$, can provide a better fit to the latest observational data. Such terms may arise in quantum gravity or effective field theory frameworks, modifying the inflationary dynamics while preserving the attractive features of the original Starobinsky model. By transforming an $f(R)$ gravity theory \footnote{See Refs.~\cite{Capozziello:2007ec,Capozziello:2011et,DeFelice:2010aj} for reviews on $f(R)$ gravity.} into its scalar-tensor representation, we systematically analyse how these corrections alter the inflaton potential and, consequently, the predictions for $n_s$, $\alpha_s$ and $r$.

Our analysis reveals that those higher-order curvature terms can adjust the inflationary observables in a way that improves consistency with ACT results, without spoiling the agreement with other datasets. This implies that the extended Starobinsky models deserve further theoretical and phenomenological scrutiny as the potential refinements to the standard paradigm of inflation.

\section{Starobinsky gravity and inflation}

A modified $f(R)$ gravity can be transformed into a scalar-tensor theory \cite{Maeda:1988ab}.  We rewrite its Lagrangian with the help of an auxiliary scalar field $\sigma$ (unless otherwise stated, we work in Planck units, $M_P=1$),
\begin{equation}
    \sqrt{-g}^{-1}\cl=f(R)~\Longrightarrow~f'(\sigma)(R-\sigma)+f(\sigma)~,
\end{equation}
such that the variation w.r.t. $\sigma$ yields the original $f(R)$ Lagrangian. If, instead, we keep $\sigma$, and transition to the Einstein frame by rescaling the metric,
\begin{equation}
    g_{\mu\nu}\rightarrow\frac{g_{\mu\nu}}{2f'(\sigma)}~,
\end{equation}
we obtain
\begin{equation}
    \sqrt{-g}^{-1}\cl=\tfrac{1}{2}R-\tfrac{1}{2}h(\sigma)(\partial\sigma)^2-V(\sigma)~,
\end{equation}
where
\begin{equation}\label{hV_of_sigma}
    h(\sigma)=\frac{3{f''}^2(\sigma)}{2{f'}^2(\sigma)}~,~~~V(\sigma)=\frac{\sigma f'(\sigma)-f(\sigma)}{4{f'}^2(\sigma)}~.
\end{equation}

Canonical scalar field $\phi$ satisfies
\begin{equation}\label{canonical_phi}
    \frac{d\phi}{d\sigma}=\sqrt{h}~,
\end{equation}
which is solved for $\phi(\sigma)$ and inverted, in order to express the scalar potential in terms of $\phi$. Even in polynomial $f(R)$ gravity, the inversion cannot always be done analytically, so one can either find $V(\phi)$ numerically, or with the help of various approximations, for example during slow-roll inflation.

Starobinsky gravity is given by
\begin{equation}\label{f_Star}
    f=\frac{1}{2}\Big(R+\frac{R^2}{6m^2}\Big)~,
\end{equation}
where $m$ is a mass parameter which can be identified with the mass of the canonical scalaron/inflaton field $\phi$. Its potential can be found by solving \eqref{canonical_phi} with \eqref{hV_of_sigma}. For convenience, from now on we will always rescale
\begin{equation}\label{sigma_rescaling}
    \sigma\rightarrow m^2\sigma~,~~~h\rightarrow h/m^4~,
\end{equation}
before solving \eqref{canonical_phi} and further calculations. After the rescaling, the solution to \eqref{canonical_phi} yields $\sigma=-3+3e^{\sqrt{2/3}\phi}$, and the potential of $\phi$ reads
\begin{equation}
    V=\frac{3}{4}m^2\big(1-e^{-\sqrt{\frac{2}{3}}\phi}\big)^2~,
\end{equation}
where the inflaton mass $m$ is fixed by the observations of the amplitude of scalar perturbations, at around $10^{-5}M_P$ \cite{Planck:2018jri}.

In order to estimate the inflationary predictions of the Starobinsky model, we introduce the potential slow-roll parameters
\begin{equation}
    \epsilon_V\equiv\frac{V_{\phi}^2}{2V^2}=\frac{V_{\sigma}^2}{2V^2h}~,~~~\eta_V\equiv\frac{V_{\phi\phi}}{V}=\frac{V_{\sigma\sigma}}{Vh}-\frac{V_{\sigma}h_{\sigma}}{2Vh^2}~,
\end{equation}
where we write them in terms of the non-canonical scalar $\sigma$, with the help of \eqref{canonical_phi} (the subscripts $\phi$ and $\sigma$ denote the respective derivatives). It is particularly useful when the solution to \eqref{canonical_phi} cannot be inverted analytically.

The number of e-folds of inflation can be estimated as
\begin{equation}\label{Delta_N_equation}
    \Delta N\simeq\int^{\phi_*}_{\phi_e}\frac{d\phi}{\sqrt{2\epsilon_V}}\simeq \int^{\sigma_*}_{\sigma_e}\frac{d\sigma\sqrt{h}}{\sqrt{2\epsilon_V}}~,
\end{equation}
where $\phi_*$ is the value of the inflaton at the horizon exit of CMB reference scale, and $\phi_e$ is the value at the end of inflation when $\epsilon_V=1$.

The inflationary observables $n_s$ and $r$ can be expressed in terms of the slow-roll parameters as
\begin{equation}\label{ns_r_eqs}
    n_s\simeq 1+2\eta_V-6\epsilon_V~,~~~r\simeq 16\epsilon_V~,
\end{equation}
calculated at the horizon exit value of the inflaton, $\phi_*$. The running of $n_s$ is given by
\begin{equation}
    \alpha_s\equiv \frac{dn_s}{d\log k}\simeq 16\epsilon_V\eta_V-24\epsilon_V^2-2\xi_V~,
\end{equation}
where $\xi_V\equiv V_{\phi}V_{\phi\phi\phi}/V^2$.

In the Starobinsky case, one can use \eqref{Delta_N_equation} to express $\phi_*$ in terms of $\Delta N$, and obtain the approximations $n_s\simeq 1-2/\Delta N$ and $r\simeq 12/\Delta N^2$. For e.g. $\Delta N=55$, we have $n_s\approx 0.964$ and $r\approx 0.004$ which were favored by the Planck data \cite{Planck:2018jri}, but are now disfavoured at $2\sigma$ by the latest ACT results \cite{ACT:2025tim}. For larger $\Delta N$, such as $\Delta N=60$, the scalar tilt is moved to the edge of the $2\sigma$ region, but still outside of $1\sigma$ region (see Figure \ref{Fig_ns_r} below). This motivates us to consider curvature corrections to $R+R^2$ gravity as a potential resolution of the tension with ACT data. In the following section we will find  $n_s$, $\alpha_s$ and $r$ in the presence of curvature corrections $R^3,R^4$, and $R^{3/2}$.

\section{Curvature corrections}

\subsection{$R^3$ correction}

First, we consider the cubic correction \cite{Huang:2013hsb,Ivanov:2021chn} 
\begin{equation}\label{f_cubic}
    f=\frac{1}{2}\Big(R+\frac{R^2}{6m^2}+\frac{\delta_3 R^3}{36m^4}\Big)~,
\end{equation}
where $\delta_3$ is the dimensionless parameter. Unlike \cite{Ivanov:2021chn} we also consider a negative  $\delta_3$.
~\footnote{A negative $\delta_3$ was excluded in  \cite{Ivanov:2021chn} because it leads to tachyonic instability at the trans-Planckian values of the scalaron  field $\phi>10$, which are irrelevant for inflation. The cubic and quartic corrections were also considered in \cite{Huang:2013hsb} in the context of Planck data. In contrast to \cite{Huang:2013hsb}, we consider a wider range of inflationary e-folds, calculate the running of $n_s$, include $R^{3/2}$ correction, and use the recent Planck+ACT constraints.}  Here the potential for the canonical scalaron can be written as
\begin{align}
    V &=3m^2(1-y)^2\frac{1+\sqrt{1+3\delta_3(y^{-1}-1)}+2\delta_3(y^{-1}-1)}{\big(1+\sqrt{1+3\delta_3(y^{-1}-1)}\big)^3}\nonumber\\
    &=\frac{3m^2}{4}(1-y)^2-\frac{3\delta_3}{8y}m^2(1-y)^3+\co(\delta_3^2)~,
\end{align}
where $y\equiv e^{-\sqrt{\frac{2}{3}}\phi}$. Since we treat the cubic term as a perturbative correction, we demand $|\delta_3|\ll 1$, so we can use it as an expansion parameter to assess its impact on $n_s$ and $r$. We will do the same for $R^4$ and $R^{3/2}$ terms as well, see below.

With the leading $\delta_3$-correction, the slow-roll parameters are given by (for $y\ll 1$)
\begin{equation}\label{SR_delta_3}
    \epsilon_V\simeq\tfrac{4}{3}y^2-\tfrac{2}{3}\delta_3,~~~\eta_V\simeq -\tfrac{4}{3}y-\tfrac{1}{3}\delta_3y^{-1}~.
\end{equation}
In Starobinsky gravity, inflation takes place for $y\ll 1$, so the main contribution to $n_s$ comes from $\eta_V$ rather than $\epsilon_V$, while the latter determines the prediction for $r$. Similarly, the leading $\delta_3$-correction comes from $\eta_V$, which implies that if we want to increase the value of $n_s$, we need $\delta_3<0$, which is also expected to increase the tensor-to-scalar ratio.

It should be mentioned that a negative $\delta_3$ also leads to a discontinuity in field space at $\sigma_{\rm dis}=2/|\delta_3|$ after the rescaling \eqref{sigma_rescaling}, where $h(\sigma_{\rm dis})=f''(\sigma_{\rm dis})=0$. The corresponding value of the canonical scalar is $e^{-\sqrt{2/3}\phi_{\rm dis}}=3|\delta_3|/(1+3|\delta_3|)$. For $|\delta_3|\ll 1$ the discontinuity is located at very large $\phi$, whereas inflation occurs for  $\phi_*\ll \phi_{\rm dis}$. Furthermore, since the $R^3$ term in the $f(R)$ formulation becomes also large, being close to the discontinuity point, the higher-order terms beyond the  $R^3$ are expected to become important in quantum gravity beyond the inflationary scale.

Though a series expansion in $\delta_3$ of the slow-roll parameters \eqref{SR_delta_3} is useful for understanding the qualitative behaviour of $n_s$ and $r$ for small $\delta_3$, in a more reasonable approach $\delta_3$ takes finite values and may not be much smaller than $y^2$ during inflation. This means that the approximation of the slow-roll parameters in 
Eq.~\eqref{SR_delta_3} may not lead to accurate predictions for the observables, and we have to take into account all orders in 
$\delta_3$ when deriving $n_s$ and $r$ at the end of this Section. The same applies to the parameters at the $R^4$ and $R^{3/2}$ terms described below. The scalar potential $V(\phi)$ and $n_s-r$ plots for the three models are shown at the end of this Section, see Figs.~\ref{Fig_potentials} and \ref{Fig_ns_r}. Figure \ref{Fig_potentials} shows that the potentials for the $R^3$-model (as well as the $R^4$-model discussed below) become steep close to the discontinuity value of $\phi$ (for negative $\delta_3$ and $\delta_4$). Having obtained numerical solutions to the inflaton equation of motion with initial conditions in the steep region at the beginning of inflation, we verified that there is no violation of slow-roll during the last $50-60$ e-folds of inflation because an inflaton velocity is suppressed by Hubble friction.

\subsection{$R^4$ correction}

Next, we consider $R^4$ term as the leading correction to Starobinsky gravity as in \cite{Huang:2013hsb,Ivanov:2021chn}, 
\begin{equation}\label{f_quartic}
    f=\frac{1}{2}\Big(R+\frac{R^2}{6m^2}+\frac{\delta_4 R^4}{48m^6}\Big)~,
\end{equation}
with the corresponding (small but negative) dimensionless parameter $\delta_4$. In the presence of the quartic term, the relevant task in \eqref{canonical_phi} amounts to solving  a cubic equation,
\begin{equation}
    \frac{\delta_4}{12}\sigma^3+\frac{1}{3}\sigma+1=\frac{1}{y}~,
\end{equation}
whose exact solution will not be needed here (see, however, \cite{Ivanov:2021chn}) because we can work directly with the non-canonical scalar $\sigma$ with the help of \eqref{hV_of_sigma} or, use the $\delta_4$-expansion for a quantitative analysis.

The slow-roll parameters with the leading $\delta_4$-corrections read ($y\ll 1$)
\begin{equation}\label{SR_delta_4}
    \epsilon_V\simeq\tfrac{4}{3}y^2-3\delta_4y^{-1},~~~\eta_V\simeq -\tfrac{4}{3}y-3\delta_4y^{-2}~.
\end{equation}
The impact of the $R^4$ term on $n_s$ and $r$ is similar to the $R^3$ term because a  negative $\delta_4$ increases the values of both $n_s$ and $r$. For negative $\delta_4$, the discontinuity is located at $\sigma_{\rm dis}^2=4/(3|\delta_4|)$, and for small enough $|\delta_4|$ one can ensure that $\phi_*\ll \phi_{\rm dis}$ for reasonable  inflation.

\subsection{$R^{3/2}$ correction}

As yet another interesting example of a fractional $R$-dependent correction, we consider adding the $R^{3/2}$ term, see e.g.,
\cite{Ketov:2010qz} where it was motivated by supergravity, 
\begin{equation}\label{f_frac}
    f=\frac{1}{2}\Big(R+\frac{\delta}{m}R^{3/2}+\frac{R^2}{6m^2}\Big)~,
\end{equation}
with the dimensionless parameter $\delta$. The approximated ($|\delta|,y\ll 1$) slow-roll parameters in this case are given by
\begin{equation}
    \epsilon_V\simeq \frac{4}{3}y^2+\frac{4\delta}{\sqrt{3}}y^{3/2}~,~~~\eta_V\simeq -\tfrac{4}{3}y-\sqrt{\frac{y}{3}}\delta~.
\end{equation}
Naively, these expressions suggest that a negative $\delta$ can reduce the value of $|\eta_V|$ and therefore increase $n_s$, as in the previous two models. Our calculations, however, show the opposite, namely, a negative $\delta$ leads to a decrease in $n_s$, whereas its positive values can lead to a small increase as is shown in Fig.~\ref{Fig_ns_r}. The reason for this behaviour is that a positive $\delta$ pushes $\phi_*$ ($\phi$ at the horizon exit) to larger values or $y_*$ to smaller values. This, in turn, leads to an overall decrease in $|\eta_V|$ (or increase in $n_s$) by raising $\delta$, but only up to a certain point. After this point, $\eta_V$ starts to increase again because the $\delta$-contribution becomes large. This effect can be seen in the turning of $n_s-r$ trajectory as we increase $\delta$ in Fig.~\ref{Fig_ns_r} on the right.

\begin{figure*}
\centering
  \centering
  \includegraphics[width=1\linewidth]{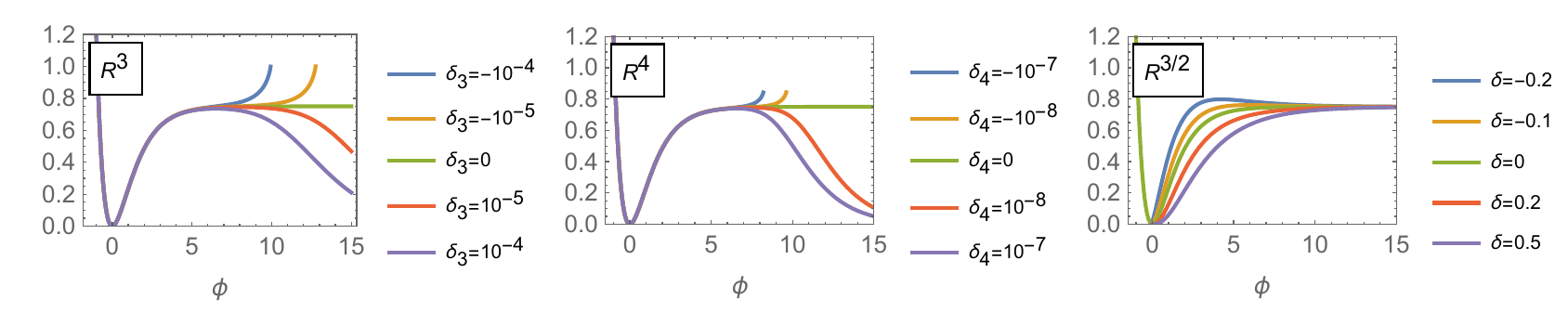}
\captionsetup{width=1\linewidth}
\caption{Scalar potential $V(\phi)/m^2$ in the three $R+R^2$ models with $R^3,R^4$, and $R^{3/2}$ corrections, respectively.}\label{Fig_potentials}
\end{figure*}

The rescaled scalar potentials $V(\phi)/m^2$ for the models with the curvature corrections $R^3,R^4$, and $R^{3/2}$ are shown in Fig.~\ref{Fig_potentials}. We vary the parameters $\delta_3,\delta_4,\delta$ between positive and negative values in order to show the changes in the potential at large $\phi$. The changes in the potentials are similar for the $R^3$ and $R^4$ cases, where positive $\delta_3$ and $\delta_4$ lead to runaway vacua at large $\phi$. The negative values lead to a sharp increase of the potential, which is stopped at the field-space discontinuity, as was mentioned above. In all our examples, inflation occurs well below the discontinuity.

In the $R^{3/2}$ model, the potential is modified at lower values of $\phi$, and coincides with the Starobinsky potential in the large $\phi$ limit. This is to be expected because  the $R^{3/2}$ is a lower-order correction than the $R^2$ and is suppressed at large $\phi$ or  large $\sigma$. For a negative $\delta$, the potential develops a local maximum, but we do consider this case because it leads to smaller $n_s$ compared to Starobinsky inflation, see \cite{Ivanov:2021chn}.

\subsection{The results for $n_s$, $r$ and $\alpha_s$}

\begin{figure*}
\centering
  \centering
  \includegraphics[width=1\linewidth]{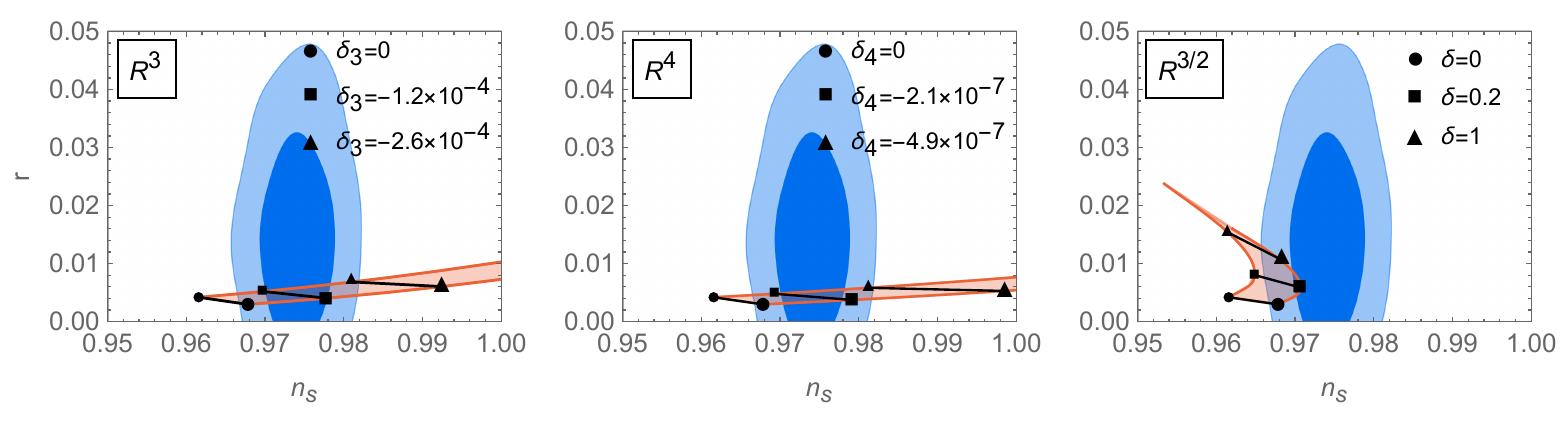}
\captionsetup{width=1\linewidth}
\caption{Scalar spectral index $n_s$ and tensor-to-scalar ratio $r$ predicted by Starobinsky inflation with $R^3,R^4$, and $R^{3/2}$ corrections. Smaller and larger marks correspond to $\Delta N=50$ and $\Delta N=60$, respectively. Blue contours represent $1\sigma$ and $2\sigma$ constraints (Planck+ACT combined) from ACT data release 6 \cite{ACT:2025tim}.}\label{Fig_ns_r}
\end{figure*}

Using the {\it exact} expressions of the scalar potential and the slow-roll parameters in our three models, we calculate the values $\phi_e$ (or $\sigma_e$) at the end of inflation ($\epsilon_V=1$), and then by inserting them in \eqref{Delta_N_equation} we find $\phi_*$ (or $\sigma_*$) as a function of $\Delta N$. Next, after fixing $\Delta N$ between $50$ and $60$, we get the scalar tilt and the tensor-to-scalar ratio from \eqref{ns_r_eqs}. The resulting plots are shown in Fig.~\ref{Fig_ns_r}, where the red trajectories correspond to gradually increasing parameters $|\delta_3|,|\delta_4|$, and $\delta$ ($\delta_3$ and $\delta_4$ are negative). Smaller and larger marks correspond to $\Delta N=50$ and $\Delta N=60$, respectively. The original Starobinsky model is shown by black circles.

\begin{figure*}
\centering
  \centering
  \includegraphics[width=1\linewidth]{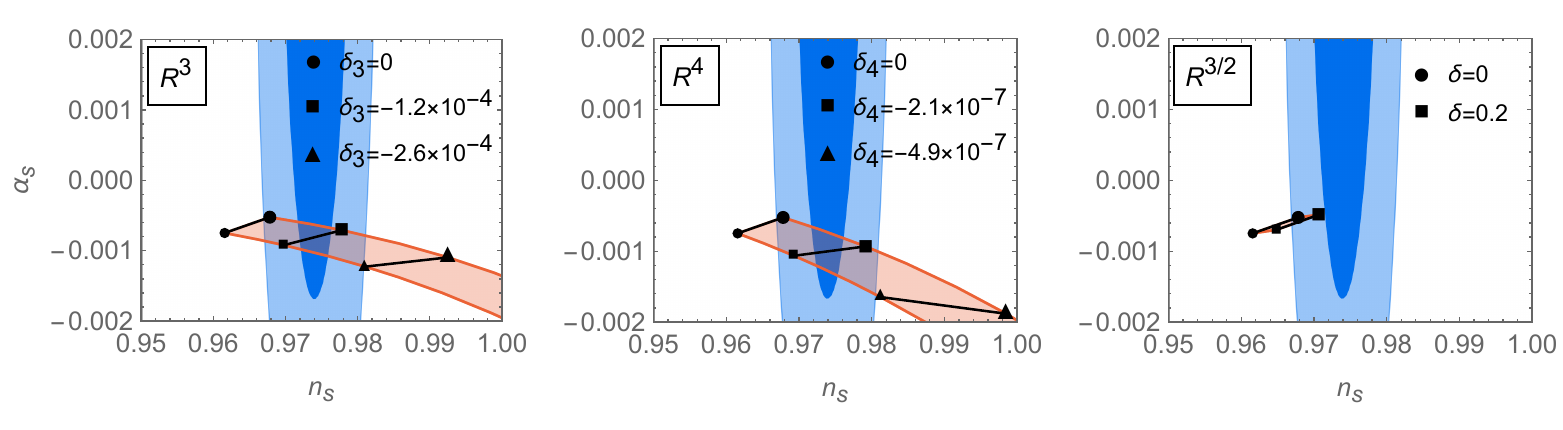}
\captionsetup{width=1\linewidth}
\caption{Running of the scalar spectral index, $\alpha_s\equiv dn_s/d\log k$, for the three corrections $R^3,R^4$, and $R^{3/2}$. The parameter choices and the notation are the same as in Fig.~\ref{Fig_ns_r}. The blue contours 
represent $1\sigma$ and $2\sigma$ constraints.}\label{Fig_nrun}
\end{figure*}

As can be seen from Fig.~\ref{Fig_ns_r}, the original Starobinsky model is outside of the $1\sigma$ region but is still within the $2\sigma$ region for $\Delta N\approx 60$. The predictions can be significantly improved if one incorporates the $R^3$ and $R^4$ corrections, or can be mildly improved with the $R^{3/2}$ correction, after suitable parameter choices. It is important to note that, in principle, one can distinguish between these three corrections as they predict different values of $r$. For example, with $\Delta N=55$ and $\delta_3=-1.19\times 10^{-4}$, we get $n_s\approx 0.974$ and $r\approx 0.0045$ for the cubic correction, and with $\delta_4=-2.02\times 10^{-7}$, we get an approximately the same $n_s$ and a slightly lower tensor-to-scalar ratio, $r\approx 0.0041$, for the quartic correction. The best fitting choice in the $R^{3/2}$ case is close to $\delta=0.2$ which, for $60$ e-folds, leads to $n_s\approx 0.9707$ and $r\approx 0.0059$.

The running of $n_s$ is shown in Fig.~\ref{Fig_nrun} with Planck+ACT constraints \cite{ACT:2025tim}. On the right-side plot ($R^{3/2}$ case), for clarity we do not show $\delta=1$ points, because they overlap with the $\delta=0$ case.
 
It is worth mentioning that contrary to Planck data, ACT favours a slightly positive value of $\alpha_{s}$, but leaves the possibility of small negative values. As with the scalar tilt itselt, our plots in Fig.~\ref{Fig_nrun} show that the $R^3$ and $R^4$ corrections can improve the predictions for the running of the tilt, to fit within $1\sigma$ region of the Planck+ACT data (dark blue area). For example, with the parameter choice $\delta_3=-1.19\times 10^{-4}$ and $\delta_4=-2.02\times 10^{-7}$ (considered above), and $\Delta N=55$, we get $\alpha_s\approx -0.0008$ and $\alpha_s\approx -0.00096$ for the cubic and quartic corrections, respectively. The fractional correction $R^{3/2}$ with $\delta=0.2$ and 60 e-folds, predicts smaller negative running $\alpha_s\approx -0.00048$ which is, however, just outside the $1\sigma$ region due to smaller $n_s$. Slightly longer inflation can bring this prediction to more favourable values by increasing $n_s$.

These results appear to favor $R^3,R^4$ models which can have $1\sigma$ compatibility with the new data, while $R^{3/2}$ can be compatible within only $2\sigma$ if inflation lasts no more than around $60$ e-folds.

\section{Conclusions}\label{Sec_concl}

We examined the impact of the curvature corrections $R^3$, $R^4$ and $R^{3/2}$ to the original Starobinsky model on the predictions of Starobinsky inflation, which was motivated by the latest ACT observations. By deriving the corresponding scalar potentials in the Einstein frame and computing the slow-roll parameters, we demonstrated that these modifications can shift the values of $n_s$ and $r$ toward a better agreement with the observational data. Though the original Starobinsky model remains viable, our results indicate that small additions to the gravitational action in the initial Jordan frame can resolve minor tensions with the high-precision ACT measurements. Other higher-derivative (perturbative) corrections, such as $R\Box R$, may also help to improve its predictions, see e.g. Ref. \cite{RomeroCastellanos:2018inv}.

It is straightforward to extend our findings to Starobinsky supergravities directly in the initial Jordan frame by using the locally supersymmetric extensions of the  $R^{3/2}$ and $R^4$ terms as well as to superstring theory in four-spacetime dimensions
by using the Grisaru-Zanon perturbative superstring correction quartic in the full curvature \cite{Toyama:2024ugg}.

All that opens new avenues for exploring modified gravity and supergravity models in the context of inflation. Future work could investigate the fundamental origin of the curvature corrections, their stability under quantum effects and their implications for reheating and primordial black hole formation, see e.g., \cite{Aldabergenov:2020bpt,Ketov:2023ykf}. The upcoming cosmological surveys CMB-S4 and LiteBIRD will provide further tests of the extended models and potentially offer deeper insights into the fundamental nature of gravity in the early universe.

\vspace{0.2cm}

{\bf Acknowledgements.} AA acknowledges the support of the National Science Foundation of China (NSFC) through the grant No.~12350410358; the Talent Scientific Research Program of College of Physics, Sichuan University, Grant 
No.~1082204112427; the Fostering Program in Disciplines Possessing Novel Features for Natural Science of Sichuan University, Grant No. 2020SCUNL209, and 1000 Talent program of Sichuan province 2021. SVK was partially supported by Tokyo Metropolitan University and the World Premier International (WPI) Research Center Initiative, MEXT, Japan. The authors are grateful to Daniel Frolovsky, Lang Liu, and Ilya Shapiro for discussions and correspondence.


\newpage

\providecommand{\href}[2]{#2}\begingroup\raggedright\endgroup

\end{document}